# ENHANCING CODE-SWITCHING ASR LEVERAGING NON-PEAKY CTC LOSS AND DEEP LANGUAGE POSTERIOR INJECTION

*Tzu-Ting Yang, Hsin-Wei Wang, Yi-Cheng Wang and Berlin Chen*

National Taiwan Normal University, Taipei, Taiwan
{tzutingyang, hsinweiwang, yichengwang, berlin}@ntnu.edu.tw

## ABSTRACT

Code-switching—where multilingual speakers alternately switch between languages during conversations—still poses significant challenges to end-to-end (E2E) automatic speech recognition (ASR) systems due to phenomena of both acoustic and semantic confusion. This issue arises because ASR systems struggle to handle the rapid alternation of languages effectively, which often leads to significant performance degradation. Our main contributions are at least threefold: First, we incorporate language identification (LID) information into several intermediate layers of the encoder, aiming to enrich output embeddings with more detailed language information. Secondly, through the novel application of language boundary alignment loss, the subsequent ASR modules are enabled to more effectively utilize the knowledge of internal language posteriors. Third, we explore the feasibility of using language posteriors to facilitate deep interaction between shared encoder and language-specific encoders. Through comprehensive experiments on the SEAME corpus, we have verified that our proposed method outperforms the prior-art method, disentangle based mixture-of-experts (D-MoE), further enhancing the acuity of the encoder to languages.

## 1. INTRODUCTION

With the widespread adoption of end-to-end (E2E) neural networks, automatic speech recognition (ASR) models have achieved significant progress and notable success across various languages. Unlike traditional ASR systems, which depend on multiple independently trained components such as acoustic and language models, E2E ASR integrates all modules into a unified neural network. This approach eliminates the inconsistencies that can occur when different components are combined. Today, over 60% of the global population is proficient in multiple languages and often unconsciously engage in code-switching (CS), alternating between different languages during daily conversations [1]. Despite many studies advancing E2E ASR to achieve human-level performance in monolingual scenarios [2][3][4][5], these systems significantly underperform when transcribing conversations that involve code-switching. This highlights the urgent need for further research into code-switching ASR [6][7][8][9].

A significant challenge when addressing code-switching in speech recognition is the scarcity of labeled data, along with substantial differences in the phonological and syntactic structures across languages [10]. When the traits of multiple languages are intertwined, E2E ASR models often struggle to effectively learn the acoustic and lexical characteristics of each language. This issue becomes even more pronounced when it involves Mandarin, which inherently contains many homophones [11]. Recent studies have mainly focused on masking heterogeneous language to establish language-aware target (LAT) [12][13], which are used to train language-independent sub-models, allowing the model to concentrate on the specific language domain. Another strategy employs a Bi-Encoder model based on the transformer architecture, where the processing of English and Mandarin is separated into two encoders, each pre-trained independently on its respective language [1][14][15][16].

While numerous studies have shown that processing different languages separately can effectively reduce semantic confusion, these methods may tend to neglect critical cross-linguistic contextual semantics. Disentangle based mixture-of-experts (D-MoE) [17] leverages the complementarity between disentanglement loss and the MoE architecture. While disentangling embeddings of different languages, D-MoE flexibly uses MoE architecture to capture the grammatical fusion patterns of languages, significantly enhancing the performance of code-switching ASR.

As a novel extension to D-MoE, this paper is dedicated to comprehensively utilizing language identification information to improve the acoustic encoder. Our main contributions are threefold: First, we inject language information into the encoder using an intermediate connectionist temporal classification (CTC) loss [18], guiding the model to learn language features more explicitly. Second, by applying language boundary alignment loss in a novel way, we can extract high time-accuracy language posteriors, making it easier for the model to infer language boundaries within frame-level sequences. Through extensive sets of experiments, we confirm that language boundaries demonstrate greater robustness compared to language peaks, thereby further enhancing the recognition performance of the ASR model. Third, we explored the feasibility of using language posteriors to facilitate deep interaction between shared encoder and language-specific encoders.

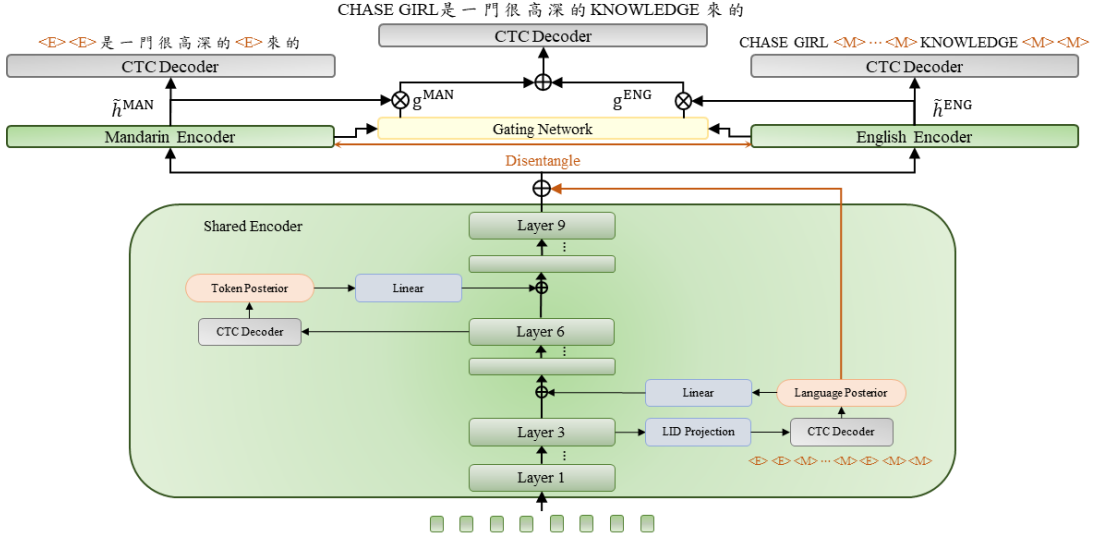

**Figure 1:** Illustration of our proposed model architecture for code-switching ASR. The symbols <M> and <E> in the figure represent the language indicator labels, <Mandarin> and <English>, respectively.

The rest of this paper is organized as follows. Section 2 will further describe our main framework, D-MoE. Section 3 elucidates our methodology, followed by Section 4 which provides a detailed analysis of the corresponding experiments. Finally, Section 5 concludes the paper and suggests future research directions.

## 2. D-MOE

D-MoE consists primarily of four components: one shared encoder, two language-specific encoders for Chinese and English respectively, and one gating network. At the outset, the shared encoder module co-processes acoustic representations, significantly reducing model parameters while effectively extracting cross-lingual acoustic information. Then, the language-specific encoders, serving as expert models using LAT, extract high-level language-dependent semantic information. Ultimately, after enhancing the distinctiveness of the outputs of expert models through disentanglement loss, D-MoE flexibly capitalizes on the MoE architecture to capture the syntactic fusion patterns of languages [14]. This approach leverages cross-lingual information while addressing the linguistic confusion caused by the lower-layer shared encoder. The total loss $\mathcal{L}_{Total}$ is represented by Equation (1):

$$\mathcal{L}_{Total} = \frac{1}{2}(\mathcal{L}_{Mix} + \mathcal{L}_{Lang}) + \lambda \cdot \mathcal{L}_{Disentangle}. \quad (1)$$

In Eq. (1), $\mathcal{L}_{Mix}$ is the CTC loss computed from the outputs of language-specific encoders, combined using the MoE architecture; $\mathcal{L}_{Lang}$ represents the CTC loss from each individual language-specific encoder; $\mathcal{L}_{Disentangle}$ is the disentanglement loss calculated using cosine distance and finally $\lambda$ is a hyperparameter that modulates these losses.

## 3. METHODLOGY

### 3.1. LID information block

Given the fact that the length $L$ of the frame-level acoustic representation $X = (x_l \in \mathbb{R}^D | l = 1, \cdots, L)$ is typically much greater than the length $S$ of the token-level ground truth sequence $\mathbf{y} = (y_s \in V | s = 1, \cdots, S)$, CTC plays a crucial role in mapping X to $\mathbf{y}$:

$$\mathcal{L}_{CTC} = -\log P(\mathbf{y}|X) = -\log \sum_{\mathbf{a} \in A} P(\mathbf{a}|X), \quad (2)$$

where the set A includes all sequences $\mathbf{a}$ that can collapse to the ground truth $\mathbf{y}$. The likelihood $\log P(\mathbf{y}|X)$ estimates the probability that the model's output features align correctly with the ground truth. The probability of each alignment $\mathbf{a}$ is given by:

$$P(\mathbf{a}|X) = \prod_{t=1}^{L} P(a_t|X), \quad (3)$$

where $a_t$ represents the frame-level posterior probability at time step $t$. Due to the nature of the backpropagation mechanism, the CTC loss imposes relatively mild constraints on the shallow encoder blocks, which may reduce the learning efficiency of the model. To address this issue, [19] integrates an intermediate loss for the encoder. Specifically, K encoder blocks are selected, and the intermediate output $X_k$ of these blocks is aligned with $\mathbf{y}$ to calculate the intermediate CTC loss $\mathcal{L}_{Inter}$:

$$\mathcal{L}_{Inter} = \frac{1}{|K|} \sum_{k=1}^{K} -\log P(\mathbf{y}|X_k). \quad (4)$$

By introducing this intermediate loss, multiple sub-models within the encoder are activated, significantly improving the learning efficiency of the shallow encoder blocks. Despite its efficiency, the frame-level independence assumption in CTC can degrade performance. To this end, self-conditioned CTC (SCCTC) enhances the correlation between frame-level acoustic features by leveraging the posterior probabilities computed from the intermediate CTC layer as a learning foundation to inform subsequent encoder blocks [20]. This method significantly enhances the robustness of frame-level acoustic features through multiple iterations. Notably, this approach shares conceptual similarities with Mask CTC [21] and Align Refine [22].

Traditional ASR encoders focus on capturing the phonetic features of speech inputs. However, compared to monolingual ASR, code-switching ASR encounters significantly more cross-language homophones. During training, the decoder relies heavily on a cross-attention mechanism to map acoustic features to the target tokens. When the encoder generates similar acoustic features for homophones, the decoder struggles to differentiate linguistic information from the encoder's output and must depend on the training text to learn language-switching patterns.

To mitigate this issue, we introduce an intermediate loss within the encoder, transforming the third encoding block of the shared encoder into an LID information block. This transformation employs a vertical multitasking learning strategy to enhance the model's sensitivity to language variations. The LID information block uses a language-only target, which includes the language indicator labels <Mandarin> and <English>, as its training target. Next, we apply the SCCTC method to the LID information block, incorporating LID posterior probabilities into the encoder's logical units to inform subsequent encoding blocks. In this way, the LID information block enhances the acoustic features with language-specific cues, enabling the encoder's output to more accurately guide the decoder in distinguishing cross-lingual homophones and thereby reducing cross-lingual semantic confusion.

### 3.2. Language boundary alignment loss

It is well known that CTC often exhibits peak behavior during training, a phenomenon attributed to the model's tendency to converge locally when starting from a uniform distribution [23]. This peakiness can cause the model to generate overly confident predictions that do not necessarily align well with the ground truth. Previously reported studies [24][25] have shown that non-peaky loss can mitigate this issue, achieving alignment capabilities comparable to current state-of-the-art (SOTA) techniques.

Although the peaky phenomenon has a limited impact on the performance degradation of ASR, in our LID information block, the posterior probability of CTC will be incorporated into the output logit. We suspect that the peakiness in these probabilities might hinder the model's ability to accurately capture frame-level language information. To address this, we propose replacing the traditional CTC loss with a non-peaky CTC loss, which provides clearer language boundaries for subsequent modules. Following [24][24][25], we modify the sequence probability evaluation in Eq. (3) to include a softmax prior $P_{prior}$, as shown below:

$$P(\mathbf{a}|X) = \prod_{t=1}^{L} \frac{P(a_t|X)}{(P_{prior})^\alpha}. \quad (5)$$

The sequence-wise softmax prior is computed from the model logits using:

$$P_{prior} = \frac{1}{L} \sum_{l=1}^{L} P(s_l|X). \quad (6)$$

Combining Eqs. (2) and (5) yields:

$$\begin{aligned} \mathcal{L}_{NPC} &= -\log \sum_{a \in A} \prod_{t=1}^{L} \frac{P(a_t|X)}{(P_{prior})^\alpha} \\ &= -\log \sum_{a \in A} exp\left(\sum_{t=1}^{L} \log(P(a_t|X)) - \alpha \log(P_{prior})\right). \end{aligned} \quad (7)$$

Introducing $P_{prior}$ increases the posterior of low-probability tokens in the alignment sequence while reducing the probability of high-probability tokens such as <Blank>, effectively reducing CTC's peakiness. This adjustment enhances the time accuracy of frame-level language posteriors. Finally, we integrate $\mathcal{L}_{Inter}$ into Eq. (1) and apply the non-peaky CTC loss to form the overall objective training function, as shown in Eq. (8):

$$\mathcal{L}_{Total} = \frac{\mathcal{L}_{NPC_{Mix}} + (\mathcal{L}_{NPC_{Lang}} + \mathcal{L}_{NPC_{Inter}})/2}{2} \\ + \lambda \cdot \mathcal{L}_{Disentangle}. \quad (8)$$

### 3.3. Deep language posterior injection

Recent studies [26][27] have introduced a two-stage decoding strategy that uses language posteriors to bias the model. Initially, they constructed an LID decoder within a multi-task learning framework. The output of this decoder was then integrated into the encoder output to achieve language biasing in the ASR model. Inspired by this strategy, we apply similar principles to the language-specific encoders within the D-MoE framework. According to [17], this approach enhances the functionality of the shared encoder, which is designed to capture cross-language information that might be missed by the language-specific encoders alone. By injecting the internal language posteriors, we enable the language-specific encoders to accurately learn language boundaries through a straightforward and effective mapping. Our method offers notable improvements over previous two-stage language biasing techniques, particularly in terms of parameter efficiency and decoding effectiveness.

**Table 1:** The SEAME corpus was analyzed in terms of the number of speakers, the total duration, and the Mandarin (MAN)-to-English (ENG) ratio.

|  | Train | Valid | Dev$_{MAN}$ | Dev$_{SGE}$ |
|---|---|---|---|---|
| Speakers | 134 | 134 | 10 | 10 |
| Duration (hr) | 97.9 | 5.2 | 7.5 | 3.9 |
| MAN/ENG | 68/32 | 67/33 | 74/26 | 37/63 |

## 4. EXPERIMENTS

### 4.1. Experimental setup

All experiments are conducted using the SEAME dataset [28], a spontaneous code-switching corpus recorded by Southeast Asian speakers. SEAME contains both intra-sentence and inter-sentence code-switching utterances. The total duration of the recorded audio is approximately 115 hours, and detailed statistics are presented in Table 1. Unlike typical code-switching in East Asia, which often involves borrowed words, code-switching in Southeast Asia is notable for the frequent switching between Mandarin and English within a single sentence. This unique linguistic behavior not only illustrates the diversity of code-switching in this region but also highlights the challenges posed by the SEAME dataset.

For our experiments, we employed Transformer CTC as the backbone architecture. In the baseline model, we configured the encoder with 15 Transformer blocks, setting the feedforward dimensions to 2048, the attention dimensions to 256, and the number of attention heads to 4. Since this paper primarily aims to enhance the encoder of the ASR model, so we used the common CTC as the decoder. To ensure a fair comparison between D-MoE method and the baseline system, we set the number of Transformer blocks contained in the shared encoder block, the Mandarin-specific encoder, and the English-specific encoder in LAE to 9, 3 and 3, respectively. To ensure adequate model capacity for handling the ground truth, we selected intermediate encoder blocks at intervals of 3 layers—specifically, layers 3, 6, 9, and 12—for computing intermediate CTC losses. Since the shared encoder in D-MoE consists of only 9 encoder blocks, we specifically targeted the ground truth as the CTC loss calculation objective in the sixth encoder block to align with the SCCTC architecture. For the language model, we used 4 Transformer blocks with dimensions identical to those of the encoder. Training transcripts were converted into the corresponding 2,624 Mandarin characters, 3,000 English BPEs, and two language indicator labels—<Mandarin> and <English>. To optimize the convergence process, we set the hyperparameter $\lambda$ in Eq. (8) to 10, ensuring a balanced scale of values within the loss. The Adam optimizer was used to perform 100 training epochs with an initial warm-up phase of 25,000 steps. The final model was obtained by averaging the top 10 checkpoints based on their validation scores. During the beam search, the beam width was set to 10. For the final evaluation, we use the mixed error rate (MER) comprising

**Table 2:** Comparison of the MER results of our method with Transformer CTC, Bi-Encoder, LAE, D-MoE, and Multi-Transformer-Transducer.

| Model | LM | MER (%) | | Params. |
|---|---|---|---|---|
| Transformer CTC | ✗ | Dev$_{MAN}$ | 21.4 | 23.05 M |
| | | Dev$_{SGE}$ | 30.3 | |
| SC-CTC | ✗ | Dev$_{MAN}$ | 20.9 | 25M |
| | | Dev$_{SGE}$ | 29.9 | |
| Bi-Encoder | ✗ | Dev$_{MAN}$ | 21.0 | 44.58 M |
| | | Dev$_{SGE}$ | 29.7 | |
| LAE | ✗ | Dev$_{MAN}$ | 21.0 | 24.46 M |
| | | Dev$_{SGE}$ | 29.5 | |
| D-MoE | ✗ | Dev$_{MAN}$ | 20.7 | 24.46M |
| | | Dev$_{SGE}$ | 29.0 | |
| Proposed Method | ✗ | Dev$_{MAN}$ | **20.0** | 27.41M |
| | | Dev$_{SGE}$ | **28.4** | |
| Multi-Transformer-Transducer | ✓ | Dev$_{MAN}$ | 20.2 | - |
| | | Dev$_{SGE}$ | 27.7 | |
| D-MoE | ✓ | Dev$_{MAN}$ | 19.2 | - |
| | | Dev$_{SGE}$ | 27.1 | |
| Proposed Method | ✓ | Dev$_{MAN}$ | **18.5** | - |
| | | Dev$_{SGE}$ | **26.3** | |

the character error rate (CER) for Mandarin and the word error rate (WER) for English.

### 4.2. Overall comparison

While both the Bi-Encoder and LAE showed significant performance improvements on the ASRU2019 [29] corpus recorded by East Asian speakers, their results on the SEAME corpus fell short of expectations. The ASRU2019 corpus contains only ~10% English content, whereas the SEAME corpus, as shown in Table 1, has at least 30% of English words. This indicates that SEAME features more diverse code-switching phenomena compared to ASRU2019, with languages interwoven at multiple levels. Such complexity may exceed the capabilities of traditional models.

We compared our proposed method with several recent and relevant modeling structures on SEAME, as detailed in Table 2. After pre-training the encoder on Aishell-1 [30] and Librispeech_clean_100 [31] to optimize the encoder, the Bi-Encoder shows modest improvements in the Dev$_{MAN}$ and Dev$_{SGE}$ test sets. Similarly, LAE, which captures cross-lingual information through shared encoder blocks, achieved a 0.2% reduction in MER on the more complex Dev$_{SGE}$ with only half the parameters. D-MoE reduces semantic confusion between different language embeddings by utilizing a complementary expert model architecture and disentanglement loss, thereby enhancing overall performance.

Our proposed method leverages the superior alignment capability of NPC loss to extract high-precision language posteriors probabilities that encapsulate language boundary concepts, enhancing the linguistic information available for subsequent modules. Compared to D-MoE, this approach results in an additional 0.5% reduction in MER across two test datasets. This confirms that although the peak phenomenon of CTC does not directly affect the performance of general ASR models, precise alignment results are crucial in the SCCTC framework. By extracting high-precision language posteriors, our method better integrates language model information, further enhancing performance. Ultimately, our proposed method achieved an impressive 18.5% MER on **Dev_MAN** and 26.3% MER on **Dev_SGE**, highlighting its effectiveness in handling the complexities of CSASR and outperforming SOTA transformer-like models.

**Table 3:** Comparison of the MER(%) across various methods of applying SCCTC.

| Model | α | Dev_MAN | Dev_SGE | Average |
|---|---|---|---|---|
| SCCTC | - | 21.1 | 30.0 | 25.55 |
| SCCTC_LIDall | - | 21.2 | 30.0 | 25.6 |
| SCCTC_LID3 | - | 20.8 | 29.7 | 25.25 |
| SCCTC_LID3 | 0.1 | 20.3 | 28.9 | 24.6 |
|  | 0.2 | **20.1** | **28.5** | **24.3** |
|  | 0.3 | **20.1** | **28.5** | **24.3** |
|  | 0.4 | 20.4 | 28.9 | 24.65 |
|  | 0.5 | 21.5 | 30.0 | 25.75 |

### 4.3. Ablation study on NPC loss

Referencing Table 3, this section focuses on the impact of deeply incorporating linguistic information and NPC on our model. Our results in the SCCTC multi-task application align closely with the results presented in [32]. By substituting the third layer's training target with LID targets, we subtly guide the model to optimal performance through the injection of language information. However, using language targets for all intermediate CTC losses results in a slight decline compared to SCCTC. This occurs because the higher-level encoder blocks, which focuses on semantic processing, benefits less from lower-level language information, leading to hindered learning.

We also examined the effect of different values of α on the model's performance. The results exhibit a U-shaped trend. When α is set to 0.2 or 0.3, NPC significantly enhances the reliability of LID posteriors with precise alignment, achieving optimal performance. Conversely, when α exceeds 0.5, the model overly amplifies low-probability tokens. This causes difficulties for the LID information block in predicting language indicator labels, resulting in notable performance degradation.

Compared to the optimal performance in Table 2, we observe that the language posterior injection discussed in Section 3.3 only reduced the MER by 0.1% on both test sets. The language posterior seems to have already enhanced the language awareness of the encoder's internal modules through a self-conditioned framework. However, if we aim to fully harness the potential for interaction with specific language encoders, simply converting the encoder's latent representations into language labels via the LID projection layer may prove inadequate.

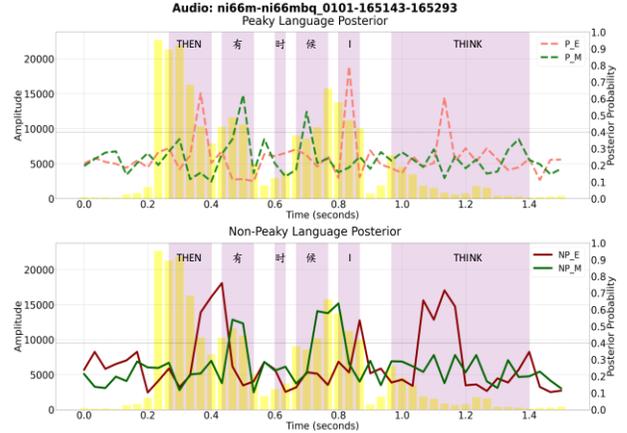

**Figure 2:** We visualized the changes in the posterior probability distributions before and after the introduction of non-peaky loss, with red and green colors representing the posterior probabilities for Chinese and English, respectively.

### 4.4. Visualizing language posteriors

Fig. 2 shows the language posterior results during CTC computation. The upper part reveals that the traditional CTC method results in spike-shaped predicted tokens. In contrast, the lower part of the figure illustrates how our model, after integrating NPC loss in a novel way, better delineates language boundaries. We also observe a latency phenomenon in the language posterior probabilities at the intermediate layers, which aligns with findings from [25] and suggests an important area for future research.

### 5. CONCLUSION

In this paper, we aim to improve the acoustic encoding of E2E ASR to address code-switching challenges. By integrating LID information into intermediate encoder layers and applying language boundary alignment loss in a novel way, we have significantly enhanced the robustness of intermediate language posteriors. For future work, we plan to pair our method with more sophisticated neural architectures and conduct in-depth investigations into a principal way to model the intricate Mandarin-English code-switching phenomena for robust ASR.

### 6. ACKNOWLEDGEMENT

This work was supported in part by E.SUN Bank under Grant Numbers 202308-NTU-02 and 202408-NTU-02. Any findings and implications in the paper do not necessarily reflect those of the sponsor.

# 7. REFERENCES

[1] S. Dalmia *et al.,* "Transformer-Transducers for Code-Switched Speech Recognition," in *Proceedings of International Conference on Acoustics, Speech and Signal Processing (ICASSP)*, 2021.

[2] Y. Zhang *et al.,* "Pushing the Limits of Semi-Supervised Learning for Automatic Speech Recognition," *arXiv preprint arXiv:2010.10504*, 2020.

[3] Q. Xu *et al.,* "Self-Training and Pre-Training are Complementary for Speech Recognition," in *Proceedings of International Conference on Acoustics, Speech and Signal Processing (ICASSP)*, 2021.

[4] Z. Gao *et al.,* "FunASR: a Fundamental End-To-End Speech Recognition Toolkit," in *Proceedings of International Speech Communication Association (INTERSPEECH)*, 2023.

[5] X. Zhou *et al.,* "MMSpeech: Multi-Modal Multi-Task Encoder-Decoder Pre-Training for Speech Recognition," *arXiv preprint arXiv:2212.00500*, 2022.

[6] N. T. Vu *et al.,* "A First Speech Recognition System for Mandarin-English Code-Switch Conversational Speech," in *Proceedings of International Conference on Acoustics, Speech and Signal Processing (ICASSP)*, 2012.

[7] K. Li *et al.,* "Towards Code-switching ASR for End-to-end CTC Models," in *Proceedings of International Conference on Acoustics, Speech and Signal Processing (ICASSP)*, 2019.

[8] Y. Peng *et al.,* "Minimum Word Error Training for Non-Autoregressive Transformer-Based Code-Switching ASR," in *Proceedings of International Conference on Acoustics, Speech and Signal Processing (ICASSP)*, 2022.

[9] B. Yan *et al.,* "Towards Zero-Shot Code-Switched Speech Recognition," in *Proceedings of International Conference on Acoustics, Speech and Signal Processing (ICASSP),* 2023.

[10] F. Zhang *et al.,* "A Study of Pronunciation Problems of English Learners in China," in *Asian Social Science*, 2009.

[11] C. -Y. Li *et al.,* "Integrating Knowledge in End-To-End Automatic Speech Recognition for Mandarin-English Codeswitching," in *Proceedings of International Conference on Asian Language Processing (IALP)*, 2019.

[12] J. Tian *et al.,* "LAE: Language-Aware Encoder for Monolingual and Multilingual ASR," *arXiv preprint arXiv:2206.02093*, 2022.

[13] T. Song *et al.,* "Language-Specific Characteristic Assistance for Code-Switching Speech Recognition," in *Proceedings of International Speech Communication Association (INTERSPEECH)*, 2022.

[14] Y. Lu *et al.,* "Bi-encoder Transformer Network for Mandarin-English Code-Switching Speech Recognition Using Mixture of Experts," in *Proceedings of International Speech Communication Association (INTERSPEECH)*, 2020.

[15] X. Zhou *et al.*, "Multi-Encoder-Decoder Transformer for Code-Switching Speech Recognition," *arXiv preprint arXiv:2006.10414*, 2020.

[16] T. Song *et al.,* "Language-Specific Characteristic Assistance for Codeswitching Speech Recognition," in *Proceedings of International Speech Communication Association (INTERSPEECH)*, 2022.

[17] T.-T. Yang *et al.,* "An Effective Mixture-of-Experts Approach for Code-Switching Speech Recognition Leveraging Encoder Disentanglement," in *Proceedings of International Conference on Acoustics, Speech and Signal Processing (ICASSP)*, 2024.

[18] J. Lee *et al.,* "Intermediate Loss Regularization for CTC-Based Speech Recognition," in *Proceedings of International Conference on Acoustics, Speech and Signal Processing (ICASSP)*, 2021.

[19] R. Sanabria *et al.,* "Hierarchical Multitask Learning with CTC," in *Proceedings of IEEE Spoken Language Technology Workshop (SLT)*, 2018.

[20] J. Nozaki *et al.,* "Relaxing the Conditional Independence Assumption Of CTC-Based ASR By Conditioning on Intermediate Predictions," in *Proceedings of International Speech Communication Association (INTERSPEECH)*, 2021.

[21] Y. Higuchi *et al.,* "Mask CTC: Non-Autoregressive End-To-End ASR with CTC and Mask Predict," in *Proceedings of International Speech Communication Association (INTERSPEECH)*, 2020.

[22] E. A. Chi *et al.,* "AlignRefine: Non-Autoregressive Speech Recognition via Iterative Realignment," in *Proceedings of the 2021 Conference of the North American Chapter of the Association for Computational Linguistics: Human Language Technologies (NAACL-HLT)*, 2021.

[23] A. Zeyer *et al.,* "Why Does CTC Result in Peaky Behavior?" *arXiv preprint arXiv:2105.14849,* 2021.

[24] R. Huang *et al.,* "Less Peaky and More Accurate CTC Forced Alignment by Label Priors," in *Proceedings of International Conference on Acoustics, Speech and Signal Processing (ICASSP)*, 2024.

[25] Z. Tian *et al.,* "Peak-First CTC: Reducing the Peak Latency of CTC Models by Applying Peak-First Regularization," in *Proceedings of International Conference on Acoustics, Speech and Signal Processing (ICASSP)*, 2023.

[26] H. Liu, *et al.*, "Reducing Language Confusion for Code-Switching Speech Recognition with Token-Level Language Diarization," in *Proceedings of International Conference on Acoustics, Speech and Signal Processing (ICASSP)*, 2023.

[27] H. Liu, *et al.*, "Enhancing Code-Switching Speech Recognition with Interactive Language Biases," in *Proceedings of International Conference on Acoustics, Speech and Signal Processing (ICASSP)*, 2024.

[28] D.-C. Lyu *et al.,* "SEAME: A Mandarin-English Code-Switching Speech Corpus in South-East Asia," in *Proceedings of International Speech Communication Association (INTERSPEECH)*, 2010.

[29] X. Shi *et al*., "The ASRU 2019 Mandarin-English Code-Switching Speech Recognition Challenge: Open Datasets, Tracks, Methods and Results," *arXiv preprint arXiv:2007.05916,* 2020.

[30] H. Bu *et al*., "AISHELL-1: An Open-Source Mandarin Speech Corpus and A Speech Recognition Baseline," in *Proceedings of the international coordinating committee on speech databases and speech I/O systems and assessment (O-COCOSDA)*, 2017.

[31] V. Panayotov *et al*., "Librispeech: An ASR Corpus Based on Public Domain Audio Books," in *Proceedings of International Conference on Acoustics, Speech and Signal Processing (ICASSP)*, 2015.

[32] L. Lonergan *et al.,* "Low-Resource Speech Recognition and Dialect Identification of Irish in A Multi-Task Framework," *arXiv preprint arXiv:2405.01293*, 2024.